\documentclass[aps,10pt,pra,twocolumn,showpacs,showkeys,groupedaddress]{revtex4-1}
\usepackage[colorlinks,linkcolor=blue,citecolor=blue,filecolor=cyan,urlcolor=blue]{hyperref}
\usepackage{amsmath}
\usepackage{graphicx}
\usepackage{bm}
\usepackage{xcolor}
\usepackage{comment}
\usepackage[normalem]{ulem}

\DeclareBoldMathCommand{\bfmu}{\mu}
\newcommand{\be}{\begin{equation}}
\newcommand{\ee}{\end{equation}}
\newcommand{\bea}{\begin{eqnarray}}
\newcommand{\eea}{\end{eqnarray}}
\def\mum{\rm\mu m}

\def\vs{{\it vs.\/}}

\def\insitu{{\it in~situ\/}}

\begin{document}

\title{Phase space tomography of cold-atom dynamics in a weakly corrugated potential}

\author{Shuyu Zhou}	
	\affiliation{Department of Physics, Ben-Gurion University of the Negev, Be'er Sheva 84105, Israel}
\author{Julien Chab\'e}
	\altaffiliation{Present address: Observatoire de la C\^ote d'Azur, Universit\'e de Nice-Sophia Antipolis, CNRS, Parc Valrose, F-06108 Nice Cedex~2, France.}
\author{Ran Salem}
	\altaffiliation{Present address: Physics Department, Nuclear Research Center Negev, Beer-Sheva 84109, Israel.}
	\affiliation{Department of Physics, Ben-Gurion University of the Negev, Be'er Sheva 84105, Israel}
\author{Tal David}
	\affiliation{Department of Physics, Ben-Gurion University of the Negev, Be'er Sheva 84105, Israel}
\author{David Groswasser}
	\affiliation{Department of Physics, Ben-Gurion University of the Negev, Be'er Sheva 84105, Israel}	
\author{Mark Keil}
	\thanks{Corresponding author}
	\email{\tt mkeil@netvision.net.il}
	\affiliation{Department of Physics, Ben-Gurion University of the Negev, Be'er Sheva 84105, Israel}	
\author{Yonathan Japha}
	\affiliation{Department of Physics, Ben-Gurion University of the Negev, Be'er Sheva 84105, Israel}	
\author{Ron Folman}
	\affiliation{Department of Physics, Ben-Gurion University of the Negev, Be'er Sheva 84105, Israel}

\begin{abstract}
We demonstrate tomographic reconstruction of the phase space distribution of atoms oscillating in a harmonic trap with weak potential corrugation caused by nanoscale imperfections in an atom chip. We find that deformations in these distributions are highly sensitive to anharmonic components of the potential. They are explained in terms of angular velocity dispersion of isoenergetic phase space trajectories. We show that the method is applicable for probing classical and quantum dynamics of cold atoms, and we note its importance for future technological applications.
\end{abstract}

\date{\today}


%?? \keywords{keywords go here} removed for ArXiv

\maketitle

\section{Introduction\label{sec:introduction}}

Weak disorder caused by corrugations in otherwise smooth or perfectly periodic potentials can have significant physical effects. It is well known, for example, that a disordered potential induces Anderson localization of quantum particles~\cite{Anderson,Nature453-891,PRL101-255702}. Small disorders can induce dramatic changes even in classical mechanics, for example, by coupling the separate dimensions so that the system becomes non-integrable~\cite{EJP32-431}. Under certain conditions, disorder can be used to control chaos, from synchronization of coupled nonlinear oscillators~\cite{Nature378-465} to stabilization of soliton solutions~\cite{PRL84-3053} and, very recently, theoretical predictions for the regularization of classical chaotic motion of cold atoms in optical lattices~\cite{PRL112-034101}. 

Potential corrugations cause density fluctuations, and even fragmentation, as ultracold atom clouds are brought close to current-carrying wires in magnetic traps~\cite{RMP79-235,PRL92-076802}. These corrugations are due to the effect of edge, surface, and bulk nanoscale defects in the wire. Solid-state structures and electronic transport characteristics can be revealed by investigating these potential corrugations~\cite{Science319-1226,PRB77-201407}. Although such potential corrugations have been reduced significantly by static techniques, such as improved wire fabrication~\cite{Folman,Reichel}, and by dynamic techniques such as using time-averaged potentials~\cite{JPB35-469,PRL98-263201}, the residual corrugation has obvious harmful effects when highly smooth magnetic potentials are needed. Similar effects are expected for permanent magnets~\cite{PRA75-063406}. In fact, any kind of engineered interaction with a nearby surface is expected to suffer from such solid-state imperfections. At a time when atom chips~\cite{AdvAtMol48-263,APB74-469} are about to be sent into space~\cite{NuclPhys243-203,CAL}, when trapped and guided matter-wave interferometers are being  designed~\cite{PRA84-033639,PRA86-043613,PRL99-060402,PRL99-173201,NatComm4-2077}, and when so-called hybrid devices are being envisioned~\cite{PRA79-040304,PRL103-043603}, investigation of such imperfections on the phase space evolution is timely. Here we study phase space evolution in a nearly harmonic trap, the most common trapping potential for cold atomic clouds. 

When an atomic cloud oscillates in a trap with a pure harmonic potential, the oscillation persists indefinitely in the absence of interatomic collisions. Even in the presence of collisions, displacement of an atomic cloud, initially at equilibrium in a harmonic trap, does not cause any change in the phase space distribution relative to the center of mass~\cite{JPB35-2383}. However, even a small potential disorder can induce damping of the oscillations~\cite{PRL98-263201} or changes in the oscillation period~\cite{PRA72-033610}.

A powerful approach for characterizing the dynamical evolution of such perturbed oscillations may be realized by measuring the phase space distribution of the atomic cloud~\cite{Nolte}. Tomography provides a method to reconstruct such phase space distributions from experimental data~\cite{Leonhardt}, as has been demonstrated for photonic states~\cite{PRL87-050402}, for single ions in very high-frequency Paul traps~\cite{PRL77-4281}, for spin states of cold atoms~\cite{Nature464-1170}, and for motional states of an atomic beam~\cite{Nature386-150,JMO44-2551}. To the best of our knowledge, however, tomography has not been applied to determine the phase space evolution of ultracold atoms in anharmonic potentials. Moreover, while groundbreaking superfluid-insulator dynamical studies have been done that also characterized the center-of-mass motion in anharmonic traps~\cite{NJP5-71}, and focusing of ultracold atoms from a box-shaped potential has been exhibited experimentally~\cite{JPB43-155002}, no real-space dynamical studies have been done at temperatures and/or densities where small nano-Kelvin disorders are expected to play a significant role~\cite{PRA81-063415,PRA88-043406}.

In this paper we demonstrate a full tomographic reconstruction of the phase space distribution of a cold atomic cloud oscillating in a harmonic trap with small potential corrugations originating in the nearby surface. We analyze these dynamics experimentally and theoretically. In addition to intrinsically interesting physical aspects of phase space evolution such as squeezing, and insights allowing better designs for technological applications, we demonstrate that even very weak potential corrugations can significantly alter the phase space distribution, thus providing a very sensitive probe of such weak corrugations.

Section~\ref{sec:experiment} of this paper outlines the cold-atom experimental setup and the results of our tomographic analysis. In~Sec.~\ref{sec:theory} we present a theoretical framework that is particularly well suited for analyzing phase space distributions as they evolve in weakly corrugated potentials, leading to an interpretation of the experimental results in~Sec.~\ref{sec:analysis}. Although our measurements are performed for a strictly classical system, Sec.~\ref{sec:quantum} discusses extensions to fully quantum systems. Finally, we summarize our results and conclusions in~Sec.~\ref{sec:summary}.

\section{Experiment\label{sec:experiment}}

\subsection{Setup and method\label{subsec:setup}}

\begin{figure}[t!]
   \vskip-\baselineskip
      \centering
      \includegraphics[width=0.33\textwidth]{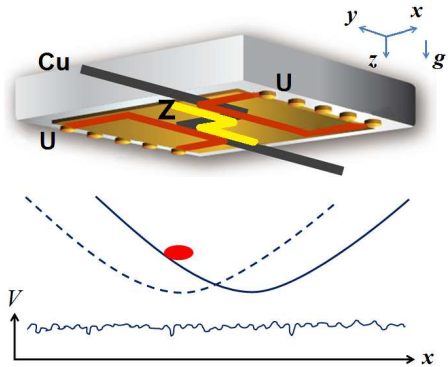}
   \vskip-0.5\baselineskip
   \caption{(Color online) (top)~The atom chip (gold surface) acts as a mirror for the magneto-optical trap and for imaging the atomic cloud. Three wires are used to generate the potentials used in this experiment. (bottom)~The trapping fields are harmonic potentials~(smooth curves) generated by the ``trapping'' wire~(Cu) whose center is~$\rm\approx\!1.2\,mm$ from the atomic cloud; the harmonic potentials may be displaced along the~$x$ axis~(solid \vs\ dashed curves) by currents through a pair of gold~U-shaped ``displacement'' wires~(U) located~$\rm320\,\mum$ from the atoms. The weak corrugation potential~(wavy line along the $x$~axis), which is highly exaggerated compared to the harmonic potentials, is imposed by a small current through the atom chip ``corrugation'' wire~(Z) located~$\rm20\,\mum$ from the atoms. External coils creating bias fields along the~$x$ and~$y$ axes are not shown.}
   \label{fig:schematic}
\end{figure}

We implement the experiment in an atom chip setup, shown schematically in~Fig.~\ref{fig:schematic}. About~$5\times10^7$ $\rm^{87}Rb$ atoms are collected by a magneto-optical trap in an ultrahigh-vacuum chamber. After molasses cooling we transfer the atomic cloud into a magnetic trap which, after~rf evaporative cooling to~$\approx\!\rm400\,nK$, has about~3000 atoms located~$\rm370\,\mum$ from the chip surface. Next, we adiabatically bring the atoms to~$\rm20\,\mum$ from the chip. The magnetic field at the trap minimum is~$\rm18\,G$, which ensures that the trap is almost perfectly harmonic in the longitudinal~($\hat x$) direction. It is also very smooth since the atoms are relatively far from the ``trapping'' wire (labeled~Cu in~Fig.~\ref{fig:schematic}). These transfers further reduce the cloud temperature to~$\rm160\,nK$; trap frequencies are~$\omega_0=\rm2\pi\times38\,Hz$ (longitudinal) and about~$\rm2\pi\times110\,Hz$ (transverse), and the elastic collision rate is~$\approx\!\rm2\,s^{-1}$. 

Currents passing in opposite directions through the two ``displacement'' wires~(labeled~U) are used to shift the trap center along the~$x$ axis. Suddenly turning off these displacement wires at the beginning of our experiments then induces oscillations of the atomic cloud along the~$x$ axis of the trap at a well-defined initial time. Experiments with currents in the ``corrugation'' wire~(labeled~Z) are conducted at a distance of~$\rm20\,\mum$; when no current flows through the corrugation wire, the distance of the atom cloud from the chip is slightly different but without any observable change in the harmonic potential. 

In order to observe the phase space distribution after an oscillation time~$t_1$, we adiabatically turn off the current in the corrugation wire and let the atoms evolve in the perfect harmonic potential for a variable time~$t_2$. This is equivalent to performing a phase space rotation~$x\to x\cos\theta-(p/m\omega_0)\sin\theta$ and~$p\to p\cos\theta+m\omega_0 x\sin\theta$, where~$\theta=-\omega_0 t_2$. We then take an~\insitu\ absorption image of the atomic density and integrate it along the transverse direction to obtain the longitudinal density~$n(x)$. The normalized density is equivalent to a projection of the phase space distribution~$P(\bar{q},\bar{p})$, where~$\bar{q}=x$ and~$\bar{p}=p/m\omega_0$ over the angle~$\theta$. The tomography algorithm, described below, reconstructs the original phase space distribution~$P(\bar{q},\bar{p})$ from a series of projections at~$0<\theta<\pi$.

Given a phase space distribution~$P(\bar{q},\bar{p})$, a projection of this distribution along an axis rotated by an angle~$\theta$ relative to the~$\bar{q}$ axis is defined as
\be 
{\rm pr}(\bar{q},\theta)=\int_{-\infty}^{\infty}d\bar{p}\,P(\bar{q}\cos\theta-\bar{p}\sin\theta, \bar{q}\sin\theta+\bar{p}\cos\theta). 
\ee

The tomography algorithm allows the reconstruction of the original phase space distribution~$P(\bar{q},\bar{p})$ from
a series of projections at~$0<\theta<\pi$. The accurate form of the mathematical transformation~\cite{Leonhardt}
\be 
P(\bar{q},\bar{p})=\frac{1}{2\pi^2}\int_0^{\pi}d\theta\int_{-\infty}^{\infty}dx K(\bar{q}\cos\theta+\bar{p}
\sin\theta-x){\rm pr}(x,\theta), \label{eq:tomography} 
\ee
with the kernel function~$K(x)=\int_0^{\infty} dk\, k\, \exp(ikx)$, is replaced by a sum over a finite number of 
projections and an approximate nondivergent form of the kernel,
\be 
K(x)=\left\{\begin{array}{ll} \frac{1}{x^2}\left[\cos k_cx+k_cx\sin k_cx-1\right] & k_cx>0.1, \\[12pt]
\frac{k_c^2}{2}\left[1-\frac{k_c^2x^2}{4}+\frac{k_c^4x^4}{72}\right] & k_cx\leq 0.1, \end{array}\right. 
\ee
where~$k_c$ corresponds to a wavelength on the order of the phase space resolution allowed by this algorithm ($k_c=\rm0.43\,\mu m^{-1}$ in our case).

In our experiment we measure the atomic density at~$t_2$ time intervals of~$\rm1\,ms$ spanning a half-period of the oscillation, giving rise to~13 images that correspond to a series of angles separated by~$\pi/13$. To improve the signal-to-noise ratio we average ten measurements for each angle.

It is helpful to note that our tomography technique allows a full reconstruction of the phase space distribution. A previous tomographic reconstruction used a free space propagation method for changing the projection angle~\cite{Nature386-150,JMO44-2551}, valid only for the restricted range~0 to~$\pi/2$, while the rest of the angles were obtained by a symmetry assumption. In contrast, our method spans the full range of~$\theta$ from~0 to~$\pi$, thereby implementing one of the procedures suggested in~Ref.~\cite{PRA78-025602}. We also note that our tomography technique can be extended to obtain the Wigner function for a quantum system under experimentally attainable conditions, as we discuss further in~Sec.~\ref{subsec:Wigner}. 

\subsection{Reconstructed phase space distributions\label{subsec:reconstruction}}

\begin{figure}[t!]
   \centering 
   \includegraphics[width=0.50\textwidth]{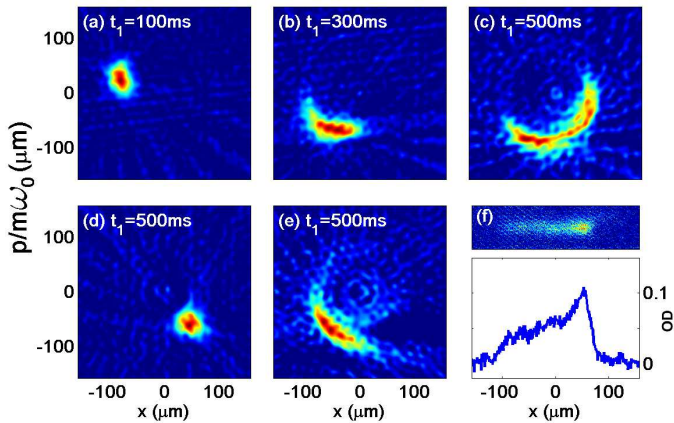}
   \caption{(Color online) (a)--(c) Reconstructed phase space distributions with a~5-mA current in the corrugation wire, an~85-$\mum$ oscillation amplitude, and oscillation times~$t_1$ as shown. The distribution in~(d) is obtained with the corrugation wire turned off, while that shown in~(e) is obtained by translating the equilibrium position of the cloud by~$\rm73\,\mum$ using the displacement wires. (f)~A typical~\insitu\ image and its corresponding one-dimensional density distribution, selected from the data set used for reconstructing the~$t_1=\rm500\,ms$ phase space distribution shown in~(c).}
   \label{fig:oscillation}
\end{figure}

In the first series of measurements~(Fig.~\ref{fig:oscillation}), we force the cloud to oscillate with an amplitude of~$\rm85\,\mum$. When there is no current in the corrugation wire, the phase space distribution maintains an approximately isotropic Gaussian shape, even after oscillating for~$\rm500\,ms$ [Fig.~\ref{fig:oscillation}(d)]. This demonstrates that the trap potential is almost purely harmonic. Applying a~$\rm5\,mA$ current to the corrugation wire causes dramatic changes. After an oscillation time of~$\rm100\,ms$, the phase space distribution develops a slight deformation, which appears more distinctly after~$\rm300\,ms$ [Figs.~\ref{fig:oscillation}(a) and~\ref{fig:oscillation}(b)]. After oscillating for~$\rm500\,ms$, the phase space distribution develops a crescent-shaped pattern totally unlike that for the purely harmonic potential [Fig.~\ref{fig:oscillation}(c)].

We conducted a second series of measurements to examine the effect produced by a different region along the corrugation potential. We displaced the equilibrium position along the longitudinal axis and forced oscillations with a slightly smaller amplitude~($\rm80\,\mum$). Differences between the resulting phase space distribution~[Fig.~\ref{fig:oscillation}(e)] and that shown in~Fig.~\ref{fig:oscillation}(c) may therefore be attributed to differences of the corrugation potential in their respective oscillation ranges.

\section{Theory\label{sec:theory}}

As a basis for understanding the experimental results, we begin with a general theoretical analysis for the phase-space evolution of a collisionless system of classical particles in a perturbed harmonic potential. Consider atoms moving in a trapping potential separable into a longitudinal part~$V(x)$ and a transverse part (along~$y$ and~$z$), such that the dynamics along~$x$ are independent of the transverse coordinates. For temperatures sufficiently in excess of the critical temperature for condensation, the evolution can be treated classically according to Newton's equations of motion~$\dot{x}=p/m$, $\dot{p}=F(x)$, where~$m$ is the mass and~$F(x)=-\partial_x V(x)$ is a position-dependent force. For a potential~$V(x)$ having a global minimum at~$x=0$, it is useful to define radial phase space coordinates~$r$ and~$\theta$ such that~$x=r\cos\theta$ and~$p=m\omega_0r\sin\theta$, with~$\omega_0$ being the harmonic frequency characterizing the potential. These equations imply that the evolution follows phase space trajectories with constant energy
\be E=\frac{p^2}{2m}+V(x)=\frac{1}{2}m\omega_0^2r^2+\Delta U(x)={\rm const}, \label{eq:E} \ee
where~$\Delta U(x)\equiv V(x)-\frac{1}{2}m\omega_0^2x^2$ is the deviation of the potential from a pure harmonic and is considered to be comparatively small.

\subsection{Angular phase space velocity dispersion\label{subsec:angular_dispersion}}

Now we calculate the equation of motion for the phase space angle~$\theta$:
\bea \dot{\theta}&=& 
\frac{1}{1+(p/m\omega_0x)^2}\left[\frac{F}{m\omega_0 x}-\frac{p^2/m}{m\omega_0 x^2}
    \right].
\eea
We write~$F=-m\omega_0^2x+\Delta F$, where~$\Delta F=-\Delta U'(x)$, and obtain
\bea \dot{\theta}&=&  
-\omega_0\left[1-\frac{x\Delta F(x)/2}
        {E-\Delta U(x)}\right],
\eea
where~$x$ can be written as $x(\theta,E)=r(\theta,E)\cos\theta$.  

The phase space distribution~$P(r,\theta,t)$ after an evolution time~$t$ in the trap is determined by the initial distribution, where each point in phase space is transformed into a new point~$r\to r(t)$ and~$\theta \to \theta(0)+\int_0^t \dot{\theta}(E,t')dt'$. If the anharmonicity of the potential is weak, then the phase space radius~$r(E,t)$ is almost constant for each energy. However, the dispersion of angular phase space velocities may cause considerable distortion of the phase space distribution. Parts of the phase space trajectories where~$x\Delta F(x)<0$ (perturbation force pointing towards the center of oscillation) accelerate the angular phase space motion, while parts with~$x\Delta F(x)>0$ (perturbation force pointing away from the center) slow it down. Phase space distortion can therefore be caused by particles moving along trajectories with different average angular velocities even though their energies may be only slightly different. 

\subsection{Oscillation period\label{subsec:period}}

Now we wish to compute the period of the motion, which is the integral over time between end points where~$\theta$ returns to its original value. This is given by
\be T_{\rm osc}=\int_0^{2\pi} \frac{d\theta}{\omega(\theta)}, \ee
where~$\omega(\theta)\equiv -d\theta/dt$. The deviation of the motion period from that of the harmonic oscillator of frequency~$\omega_0$ is given by
\be \delta T_{\rm osc}(E)=\frac{1}{\omega_0}\int_0^{2\pi}d\theta\ \frac{x\Delta F/2}{E-\Delta U-x\Delta F/2}. 
\label{eq:dT} \ee
This is true whenever the denominator inside the integral is positive. Otherwise, the trajectory of phase space motion
would not complete a full round trip around the origin~$x=p=0$.

In the limit of a weak perturbation, such that~$\Delta U/E\ll 1$ and~$x\Delta F/E\ll 1$, we may neglect the position-dependent terms in the denominator and make a coordinate transformation
\be d\theta=\frac{dx}{dx/d\theta}=\frac{dx}{(dr/d\theta)\cos\theta-r\sin\theta}. \ee
By taking the derivative of~Eq.~(\ref{eq:E}) with respect to~$\theta$ it can be shown that
\be \frac{\partial r}{\partial\theta}=r\tan\theta\ \frac{x\Delta F/2}{E-\Delta U-x\Delta F/2}, \ee
such that 
\be d\theta\approx\ -\frac{dx}{r\sin\theta}\ \frac{1}{1-x\Delta F/2E}. \ee
The second term in the denominator of the second factor can therefore be neglected in the limit~$x\Delta F\ll E$. 
By using~$r\sin\theta=\pm \sqrt{r^2-x^2}$ and taking the lowest-order approximation for~$r$, namely,~$r\approx x_{\rm max}(E)=\sqrt{2E/m\omega_0^2}$, where~$x_{\rm max}$ is the distance of the turning point in the unperturbed harmonic potential from the center, we then obtain the lowest-order expression for the period change due to the perturbation
\be \delta T_{\rm osc}(E)\approx \frac{1}{\omega_0E}\int_{-x_{\rm max}}^{x_{\rm max}} \frac{x\, \Delta F(x)\, dx}
{\sqrt{x_{\rm max}^2-x^2}}. \label{eq:dTapprox} \ee 

Equation~(\ref{eq:dTapprox}) implies that the main contribution to the change in the period of motion for a given energy~$E$ comes from the corrugated potential gradient near the classical turning points. The sign of the contribution depends on the direction of the force~$\Delta F(x)$ as discussed above.  

\section{Interpretation and analysis\label{sec:analysis}}

We now use the preceding analysis to predict the phase space evolution for a classical system that is weakly perturbed by an anharmonic potential and to compare with the experimental results of~Fig.~\ref{fig:oscillation}. 

\begin{figure}[b!]
      \includegraphics[width=0.45\textwidth]{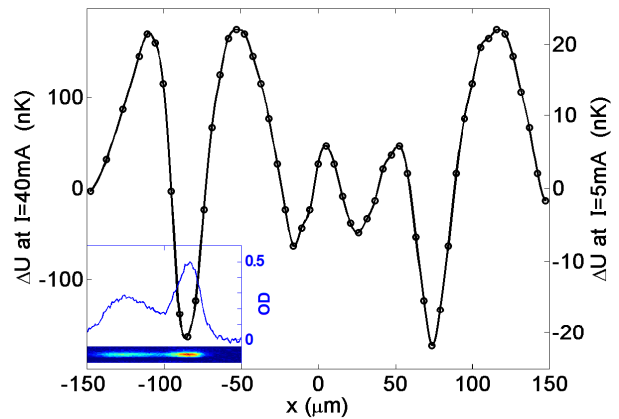}
      \vskip-0.5\baselineskip
   \caption{(Color online) The potential corrugation at a distance of~$\rm20\,\mum$ from the corrugation wire~(Z in~Fig.~\ref{fig:schematic}). The potential is shown for a corrugation-wire current of~$\rm40\,mA$ as reconstructed from a series of measurements, like those in the inset, that overlap along the~$x$ axis. The right-hand scale shows the potential as it would correspond to the corrugation-wire current of~$\rm5\,mA$ used in the oscillation experiments. The inset shows a typical image of the atomic cloud optical density~(OD) distribution for one particular current in the displacement wires.} 
   \label{fig:corrugation}
\end{figure}

\subsection{Potential corrugations\label{subsec:corrugation}}

In our case, the potential corrugation~$\Delta U(x)$ was extracted from a series of measurements of the equilibrium density~$n(x)$ of the atomic cloud in the presence of a harmonic potential~\cite{PRL98-263201} centered~$\rm20\,\mum$ from multiple positions along the corrugation wire~(Fig.~\ref{fig:corrugation}). In order to observe atom density modulations due to current fluctuations more clearly, we increased the corrugation-wire current to~$\rm40\,mA$, an eightfold increase over that used in our experiments. The trapping-wire current was correspondingly reduced to maintain the atom cloud at a distance of~$\rm20\,\mum$ from the corrugation wire. As demonstrated in previous work~\cite{PRB77-201407}, variations in the magnetic field at this distance are affected only by imperfections in the wire edge over a scale longer than~$\approx\!\rm20\,\mum$. 

In particular, the main features of the corrugation potential are two wells with amplitudes of~$\pm\rm22\,nK$ centered at~$x\rm\approx\!\pm80\,\mum$ caused by mismatches at the edges of the field of view of the lithography process, as detailed in the~Appendix. In between lie two shallower wells of amplitude~$\pm\rm7\,nK$. These potential corrugation amplitudes are shown on the right-hand scale of~Fig.~\ref{fig:corrugation}, corresponding to the~5-mA current used in the oscillation experiments.

\begin{figure}[b!]
   \includegraphics[width=0.45\textwidth]{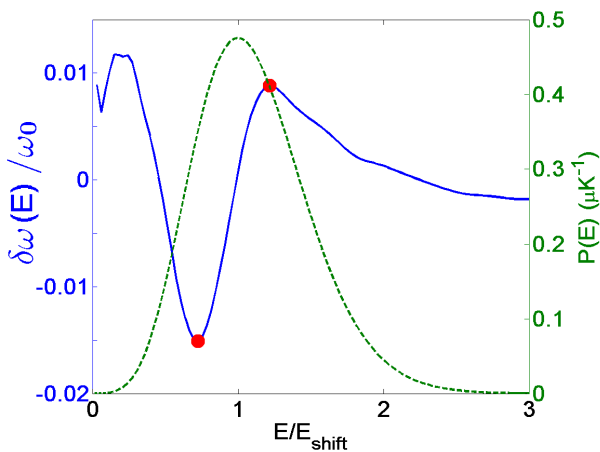}
   \vskip-0.5\baselineskip
   \caption{(Color online) Frequency shift~$\delta\omega$~(solid curve) as a function of the longitudinal energy of an atom in a harmonic trap centered at~$x=0$ in the presence of the corrugation potential shown in~Fig.~\ref{fig:corrugation} for a current of~$\rm5\,mA$. The energy is in units of~$E_{\rm shift}$ (see text). Energy distribution~(dashed curve) for an atomic ensemble prepared in the initial trap centered at~$x_{\rm shift}$. The ensemble contains atoms whose energies span oscillation frequencies ~$\delta\omega/\omega_0\approx-0.015$ to~$\delta\omega/\omega_0\approx+0.009$~(red dots). The evolution that follows leads to the simulated phase space distributions shown in~Fig.~\ref{fig:simulation}(a).}
   \label{fig:freq_dispersion}
\end{figure}

\subsection{Analysis\label{subsec:analysis}}

In our first experiment~[Figs.~\ref{fig:oscillation}(a)--~\ref{fig:oscillation}(c)] the two classical turning points of the oscillation lie near the two deepest wells of the corrugation potential (shown in Fig.~\ref{fig:corrugation}). This implies a force in the outward direction for lower energies, giving rise to a longer oscillation period~[$\delta T_{\rm osc}>0$ in~Eq.~(\ref{eq:dTapprox})] and hence a smaller oscillation frequency, while higher energies correspond to a force directed inward and hence a higher oscillation frequency. Conversely, in our second experiment the two turning points are located at~$x\approx0$ and~$x\rm\approx160\,\mum$, where the potential wells are much shallower. Qualitatively, this should give rise to a smaller phase space angular dispersion, as shown by~Fig.~\ref{fig:oscillation}(e).

For a more quantitative comparison, we first calculate the oscillation frequency for any given particle energy, as shown in~Fig.~\ref{fig:freq_dispersion}. Preparing a random Boltzmann distribution of atoms in a trap shifted by~$x_{\rm shift}=\rm85\,\mum$ and then releasing it suddenly to the original position, as performed in the experiments, produces an energy distribution centered at~$E_{\rm shift}=\frac{1}{2}m\omega_0^2x_{\rm shift}^2$ and having a standard deviation~$\Delta E\approx \sqrt{k_BT(k_BT+2E_{\rm shift})}$. This energy distribution (shown in~Fig.~\ref{fig:freq_dispersion}) involves phase space trajectories with frequency shifts between~$\delta\omega/\omega_0\approx-0.015$ and~$\delta\omega/\omega_0\approx+0.009$ (red dots in~Fig.~\ref{fig:freq_dispersion}). It follows that the angular dispersion after time~$t_1$ is~$\Delta\theta\sim0.024\omega_0 t_1\approx2.85$~rad, as indeed observed in the experimental phase space distribution~[Fig.~\ref{fig:oscillation}(c)].

\subsection{Simulations of the classical system\label{subsec:simulation}}

\begin{figure}[t!]
      \centering\hspace{-4mm}
      \includegraphics[width=0.50\textwidth]{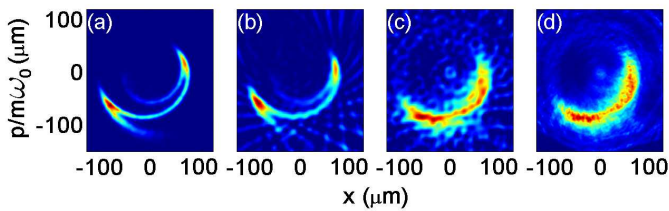}	
   \vskip-0.5\baselineskip
   \caption{(Color online) Phase space distributions obtained by simulation and experiment for oscillation times~$t_1=\rm500\,ms$. (a)~The simulation was based on preparing an ensemble of atoms with a thermal distribution in the initial trap and then adding a proper phase space angle according to~$\delta\omega(E)$ in~Fig.~\ref{fig:freq_dispersion}. (b)~Newton's equations of motion were solved numerically for each atom, and collisions were included. We also repeated the tomography algorithm and took into account the finite resolution of the optical system. The similarity between~(a) and~(b) shows that for a collision rate of about~$\rm2\,s^{-1}$ there is no significant smearing of the phase space distribution, even after~$\rm500\,ms$. (c)~The experimental results [repeated from Fig.~\ref{fig:oscillation}(c)] are explained to a high degree by the theory~[(a) and~(b)]. (d)~An alternative maximum-likelihood tomography algorithm~\cite{EPL59-694} gives similar results.}
   \label{fig:simulation}
\end{figure}

In Fig.~\ref{fig:simulation}(a) we show the results of a simulation based on adding a proper phase space angle to each atom in the random distribution according to~$\delta\omega(E)$ in~Fig.~\ref{fig:freq_dispersion}. For comparison,~Fig.~\ref{fig:simulation}(b) shows the phase space distribution resulting from a full calculation based on a direct numerical solution of Newton's equations of motion in the presence of the corrugation potential. In addition, we account for collisions by assuming that the probability of collision between each pair of atoms in a small volume~$V$ during a time interval~$\tau$ is~$\sigma_{\rm el}v_{\rm rel}\tau/V$, where~$\sigma_{\rm el}=\rm8\times10^{-12}\,cm^2$ is the~$s$-wave collision cross section and~$v_{\rm rel}$ is the relative velocity between the two atoms. This implies that higher collision probabilities are expected between pairs of atoms which occupy the same volume but have opposite velocity. Such events become more probable when the phase space angular distribution becomes more dispersed. When a collision occurs, the two atoms scatter into new velocities along the~$x$ direction, and the center of mass velocity is conserved, while the energy is redistributed between the longitudinal and transverse degrees of freedom. Comparing these two simulations suggests that the collisionless phase space features are not significantly smeared in the presence of a small number of collisions (up to one collision per atom).

Under conditions where collisions are rare and dimensions are separable, the phase space area is conserved during the cloud evolution, and lower densities and temperatures of the atomic ensemble can be reached, as in $\delta$-kick cooling~\cite{PRL78-2088}. 

\section{Extension to a quantum state\label{sec:quantum}}

Our experiment was performed at a temperature for which the ensemble of cold atoms behaves purely classically. Here we demonstrate explicitly, for realistic conditions, that our tomographic reconstruction procedure is also applicable to a fully quantum situation. In addition, we show that the stretching of an initially localized classical phase-space distribution, caused in our experiment by the presence of anharmonic perturbations, is analogous to a quantum squeezing effect whose occurrence we demonstrate for an anharmonic trap when the initial distribution is a minimum uncertainty distribution.

\subsection{Phase space tomography for the quantum case\label{subsec:Wigner}}

In order to demonstrate the applicability of the tomographic method for reconstructing a quantum phase space distribution (the Wigner function), we consider the simple situation of a Bose-Einstein condensate~(BEC) of noninteracting atoms in a harmonic trap. Such a~BEC is achievable by magnetically tuning the scattering length to a small positive value, e.g.,~by using a magnetic field of about~$\rm165.6\,G$ for~$\rm^{85}Rb$~\cite{PRL85-1795,PRA67-060701,RSI81-063103}. 

In our example, the atoms are initially in the ground state of a harmonic potential whose frequency is~$\omega=2\pi\times\rm38\,Hz$, as in our experiments. We envision a light pulse or a magnetic gradient pulse combined with microwave pulses~\cite{NatComm4-2424} that transfers the atoms into a superposition of two momenta~$\pm k=\pm 2\pi/\rm2\,\mum$, such that the wave function has the form
\begin{equation}
\psi(x,t=0)=\frac{e^{-x^2/2\sigma^2}}{\sqrt{2\pi}\sigma}\left(e^{ikx}+e^{-ikx}\right),
\end{equation}
where~$\sigma=\sqrt{\hbar/m\omega}=\rm1.76\,\mum$ is the initial width of the Gaussian wave packet. 

\begin{figure}[t!]
      \centering
      \includegraphics[width=0.45\textwidth]{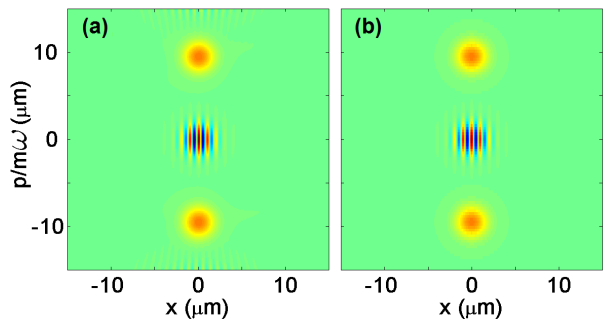}
   \caption{(Color online) Phase space distribution for a trapped noninteracting~BEC split into a superposition of two momenta. (a)~Wigner function reconstructed by keeping the~BEC in the harmonic potential for~times~$t_j=(j-1)\pi/\omega, j=1,\dots,60$ and then allowing free expansion for a flight time of~$t_f=\rm30\,ms$. (b)~Wigner function calculated directly from the wave function~[Eq.~(\ref{eq:W})]. The green background color represents values close to zero, while red is positive and blue is negative.}
   \label{fig:Wigner}
\end{figure}

The Wigner function of such a~BEC may be obtained tomographically by letting it evolve in the harmonic trap for multiple times~$t_j$, which correspond to phase space rotation angles~$\theta_j=-\omega t_j$. However, the density distribution for our~BEC varies in space with a period of~$\rm1\,\mum$, which is not resolvable with the optical imaging of our experiment. In order to image the atoms with a practical resolution we turn off the harmonic trap after each time~$t_j$ and let the atoms expand freely for an additional time~$t_f$. After this time of flight, each point in phase space transforms as
\begin{equation}
x \to x+\frac{p}{m}t=\frac{1}{\cos\theta_f}\left[x\cos\theta_f-\frac{p}{m\omega}\sin\theta_f\right] 
\end{equation}
and~$p\to p$, where~$\theta_f=-\tan^{-1}(\omega t_f)$. Since the spatial density distribution is obtained by integrating over the coordinate~$p$, it follows that the resulting spatial density after expansion is the same as if phase space is rotated by an angle~$\theta_f$ and the $x$ coordinate is then stretched by a factor~$[\cos\theta_f]^{-1}=\sqrt{1+\omega^2t_f^2}$. For~$t_f=\rm30\,ms$, this ``stretching factor'' is~$7.2$, which is large enough to provide sufficient imaging resolution for reconstructing the initial phase-space distribution.

In Fig.~\ref{fig:Wigner} we present the reconstructed Wigner function obtained with this method. We compare it to the exact Wigner function, which is obtained directly from the wave function~$\psi(x)$,
\begin{equation}
\hspace{0mm}W(x,p)=\frac{1}{2\pi\hbar}\int_{-\infty}^{\infty}\!\!d\eta\ \psi^*\!\left(x\!-\!\frac{\eta}{2}\right) \psi\!\left(x\!+\!\frac{\eta}{2}\right)e^{-ip\eta/\hbar},\hspace{-2mm}
\label{eq:W}
\end{equation}
and we conclude that the tomographic reconstruction method applied in this paper can accurately be extended to the quantum case.

\subsection{Quantum squeezing in an anharmonic trap\label{subsec:squeezing}}

\begin{figure}[t!]
      \centering
      \includegraphics[width=0.45\textwidth]{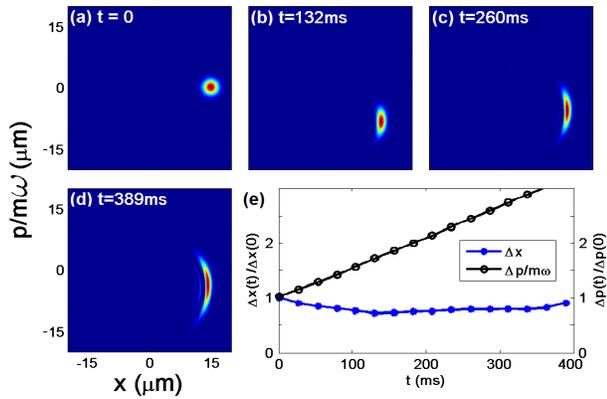}
   \caption{(Color online) Effect of a small quartic anharmonicity on the phase space evolution of a trapped single-particle quantum system. (a)--(d)~Evolution of the phase space density for different times in the trap. (e)~Uncertainties in position~(blue line) and momentum~(black line, expressed as~$p/m\omega$) as a function of time; note that the minimum position uncertainty occurring near~$t=\rm130\,ms$ is~$\approx\!30\%$ below the minimum-uncertainty initial state.}
   \label{fig:squeezing}
\end{figure}

Here we wish to demonstrate that the evolution of a quantum wave packet in a harmonic trap with an anharmonic perturbation leads to squeezing in a manner similar to the stretching of the classical phase space distribution observed in our experiment. We consider an anharmonic potential of the form
\begin{equation} V(x)=\tfrac{1}{2}m\omega^2x^2\left(1+x^2/w^2\right),
\end{equation}
where the harmonic frequency~$\omega$ is chosen to be~$2\pi\times\rm38\,Hz$ as above and the quartic term is 
characterized by the distance~$w=\rm100\,\mum$ at which the quartic addition is equal to the harmonic contribution. 
We start from a noninteracting~BEC (equivalent to a single particle) in the ground state of the harmonic potential and 
shift the potential by~$\rm15\,\mum$ from its initial center. In Figs.~\ref{fig:squeezing}(a)--\ref{fig:squeezing}(d) we show the phase space distribution as the~BEC evolves in the anharmonic trap. In Fig.~\ref{fig:squeezing}(e) we present the position and momentum uncertainties~$\Delta x$ and~$\Delta p/m\omega$ as a function of evolution time, a few milliseconds after the wave-packet center passes the initial turning point. Note that the anharmonicity also increases the oscillation period by about~$1.8\%$.

For the parameters chosen in this demonstration the maximum squeezing (minimum~$\Delta x$) is achieved after about~$\rm130\,ms$~(five oscillations in the trap). After this time the phase space distribution becomes progressively narrower in the radial direction~[Fig.~\ref{fig:squeezing}(d)], but the  position uncertainty begins to grow. The stretching seen along the arc of this quantum phase-space evolution is similar to that seen for our experiment with classical particles~(Fig.~\ref{fig:simulation}) and is caused by the small potential anharmonicity in both cases.

\section{Summary and conclusions\label{sec:summary}}\vskip-\baselineskip 

We have studied the classical phase space evolution of trapped atoms oscillating in a harmonic potential with static corrugations along one dimension. For oscillation times on the order of the mean time between collisions, phase space variables propagate along nearly circular iso-energetic trajectories whose angular velocity dispersion is determined mainly by potential corrugations near the classical turning points. We observe deformations of the phase space distribution that are sensitive to fine details of the corrugations, and we show that these classical deformations are analogous to squeezing in quantum systems.

We have demonstrated a phase space tomography method that can serve as a tool for probing the dynamics of trapped atoms. The method is sensitive to potential corrugations having an~rms amplitude of~$\rm\sim\!10\,nK$. In view of~Eq.~(\ref{eq:dTapprox}), this sensitivity may be improved by reducing the oscillation frequency and amplitude. Further improvements may be achieved by probing a longer time evolution of cold fermions or a one-dimensional ultracold Bose gas~\cite{Nature440-900,NJP12-055023}, where the effects of collisions are suppressed. If the phase space distribution were generalized to a Wigner function~\cite{Leonhardt,JOptB5-420}, the method could also be applied to coherent matter waves~\cite{PRA81-065602} or to quantum dynamics~\cite{PRA78-025602,PRA55-2109}.

Probing, understanding, and controlling the phase space evolution of cold atoms in the presence of a corrugated potential is important, for example, for future schemes of guided matter-wave interferometry~\cite{PRL99-060402,PRL99-173201} and free-oscillation atom interferometry~\cite{PRA84-033639,PRA86-043613}. Quantum phase dispersion and the visibility of interference in such interferometry would be determined by the propagation dynamics along separated paths, whose general features are expected to be similar to those of the classical evolution described above. Applying tomography as described in this paper may be done either by trapping the wave packet at the output port of the interferometer or by mapping the interferometer loop onto a phase space loop. Such studies may shed new light on the connection between classical and quantum treatments of dephasing~\cite{PRA41-3436}, with the present study focusing on static fluctuations. 

\begin{acknowledgments}\vskip-\baselineskip

We are grateful to the members of the atom chip group and especially to S.~Machluf for helpful discussions and to Z.~Binshtok, Y.~Bar-Haim, B.~Hadad, and J.~Jopp for technical support. We also thank the Ben-Gurion University fabrication facility for providing the atom chip and the~BGU machine shop team for constructing much of the apparatus. This work was supported by the Israeli Science Foundation and by the~FP7 European consortium ``matter-wave interferometry''~(601180). We also acknowledge support from the~PBC program for outstanding postdoctoral researchers of the Israeli Council for Higher Education and from the Ministry of Immigrant Absorption (Israel).

\end{acknowledgments}

\appendix*\section{Quantifying the potential corrugations\label{sec:appendix}}

\begin{figure}[h!]
      \centering
      \includegraphics[width=0.45\textwidth]{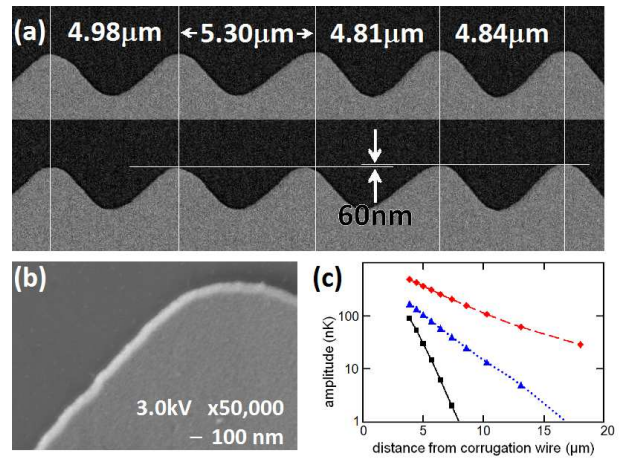}
   \caption{(Color online) (a)~Scanning-electron-microscope images showing one edge of the corrugation wire~(Z in~Fig.~\ref{fig:schematic}). The lower scan is a copy of the upper scan, displaced one period to the right. The vertical white lines are drawn through the peaks of the upper scan and show that the second period is~$\approx\!10\%$ longer than the other periods due to a fabrication error occurring every~$\rm160\,\mum$ along the wire. An additional imperfection, almost imperceptible in the image, is also indicated as a transverse shift of~$\approx\!\rm60\,nm$. (b)~Close-up view showing that random edge imperfections are~$\lesssim\!\rm20\,nm$. The wire is~$\rm8\,\mum$ wide and~$\rm0.5\,\mum$ thick. (c)~Corrugation potential amplitudes as a function of distance from the corrugation wire, as calculated for the~5-$\mum$ edge modulation~(solid black curve) and for the~160-$\mum$-period fabrication errors exhibited in~(a):~0.5-$\mum$ longitudinal stretch~(blue dotted curve) and~60-nm transverse shift~(red dashed curve). The deepest corrugations observed in the experimental potential~(Fig.~\ref{fig:corrugation}) are quantitatively reproduced by the latter curve at a distance of~$\rm20\,\mum$.}
   \label{fig:stitch}
\end{figure}

For the purpose of a different experiment which is not reported in this paper, the edges of the corrugation wire are fabricated with a modulation period of~$\rm5\,\mum$, as shown in~Fig.~\ref{fig:stitch}, giving rise to a short-range sinusoidal potential. As demonstrated in previous work~\cite{PRB77-201407}, however, the magnetic field for distances~$\ge\!\rm20\,\mum$ used in the oscillation experiments of this study is not influenced by the~5-$\mum$ periodic edge modulation, and variations of the magnetic field are dominated by imperfections in the wire edge over a longer scale. Explicit numerical calculations of the magnetic potential confirm that residual sinusoidal corrugation [due to the~5-$\mum$-period modulation of the corrugation-wire edges; solid curve in~Fig.~\ref{fig:stitch}(c)] is~$\ll\rm1\,nK$ and is therefore negligible under the conditions of the oscillation experiments. Conversely, the corrugation wire fabrication process introduced slight mismatches at the edges of the lithographic field of view with a period of~$\rm160\,\mum$. Specifically, a transverse shift of the corrugation wire by about~$\rm60\,nm$ produces a corrugation amplitude of about~$\rm20\,nK$ [dashed curve in~Fig.~\ref{fig:stitch}(c)], observed experimentally as the two deepest potential wells at~$x=\pm\rm80\,\mum$ in~Fig.~\ref{fig:corrugation}.\vfill\eject

\bibliography{phase_space}

\begin{thebibliography}{10}

\bibitem{Anderson}
P.~W. Anderson.
\newblock {Absence of Diffusion in Certain Random Lattices}.
\newblock {\em Phys. Rev.}, 109:1492, 1958.
\newblock \href{http://prola.aps.org/pdf/PR/v109/i5/p1492_1}{\tt
  doi=10.1103/PhysRev.109.1492}.

\bibitem{Nature453-891}
J.~Billy, V.~Josse, Z.~Zuo, A.~Bernard, B.~Hambrecht, P.~Lugan, D.~Cl\'ement,
  L.~Sanchez-Palencia, P.~Bouyer, and A.~Aspect.
\newblock {Direct observation of Anderson localization of matter waves in a
  controlled disorder}.
\newblock {\em Nature (London)}, 453:891, 2008.
\newblock
  \href{http://www.nature.com/nature/journal/v453/n7197/pdf/nature07000.pdf}{\tt
  doi=10.1038/nature07000}.

\bibitem{PRL101-255702}
J.~Chab\'e, G.~Lemari\'e, B.~Gr\'emaud, D.~Delande, P.~Szriftgiser, and J.~C.
  Garreau.
\newblock {Experimental Observation of the Anderson Metal-Insulator Transition
  with Atomic Matter Waves}.
\newblock {\em Phys. Rev. Lett.}, 101:255702, 2008.
\newblock \href{http://prl.aps.org/pdf/PRL/v101/i25/e255702}{\tt
  doi=10.1103/PhysRevLett.101.255702}.

\bibitem{EJP32-431}
J.~Masoliver and A.~Ros.
\newblock {Integrability and chaos: the classical uncertainty}.
\newblock {\em Eur.J.Phys.}, 32:431, 2011.
\newblock
  \href{http://iopscience.iop.org/0143-0807/32/2/016/pdf/0143-0807_32_2_016.pdf}{\tt
  doi=10.1088/0143-0807/32/2/016}.

\bibitem{Nature378-465}
Y.~Braiman, J.~F. Lindner, and W.~L. Ditto.
\newblock {Taming spatiotemporal chaos with disorder}.
\newblock {\em Nature (London)}, 378:465, 1995.
\newblock
  \href{http://www.nature.com/nature/journal/v378/n6556/pdf/378465a0.pdf}{\tt
  doi=10.1038/378465a0}.

\bibitem{PRL84-3053}
N.~V. Alexeeva, I.~V. Barashenkov, and G.~P. Tsironis.
\newblock {Impurity-Induced Stabilization of Solitons in Arrays of
  Parametrically Driven Nonlinear Oscillators}.
\newblock {\em Phys. Rev. Lett.}, 84:3053, 2000.
\newblock \href{http://prl.aps.org/pdf/PRL/v84/i14/p3053_1}{\tt
  doi=10.1103/PhysRevLett.84.3053}.

\bibitem{PRL112-034101}
T.~Wulf, B.~Liebchen, and P.~Schmelcher.
\newblock {Disorder Induced Regular Dynamics in Oscillating Lattices}.
\newblock {\em Phys. Rev. Lett.}, 112:034101, 2014.
\newblock
  \href{http://journals.aps.org/prl/pdf/10.1103/PhysRevLett.112.034101}{\tt
  doi=10.1103/PhysRevLett.112.034101}.

\bibitem{RMP79-235}
J.~Fort\'agh and C.~Zimmermann.
\newblock {Magnetic microtraps for ultracold atoms}.
\newblock {\em Rev. Mod. Phys.}, 79:235, 2007.
\newblock \href{http://journals.aps.org/rmp/pdf/10.1103/RevModPhys.79.235}{\tt
  doi=10.1103/RevModPhys.79.235}.

\bibitem{PRL92-076802}
D.-W. Wang, M.~D. Lukin, and E.~Demler.
\newblock {Disordered Bose-Einstein Condensates in Quasi-One-Dimensional
  Magnetic Microtraps}.
\newblock {\em Phys. Rev. Lett.}, 92:076802, 2004.
\newblock \href{http://prl.aps.org/pdf/PRL/v92/i7/e076802}{\tt
  doi=10.1103/PhysRevLett.92.076802}.

\bibitem{Science319-1226}
S.~Aigner, L.~{Della Pietra}, Y.~Japha, O.~Entin-Wohlman, T.~David, R.~Salem,
  R.~Folman, and J.~Schmiedmayer.
\newblock {Long-Range Order in Electronic Transport Through Disordered Metal
  Films}.
\newblock {\em Science}, 319:1226, 2008.
\newblock \href{http://www.sciencemag.org/content/319/5867/1226.full.pdf}{\tt
  doi=10.1126/science.1152458}.

\bibitem{PRB77-201407}
Y.~Japha, O.~Entin-Wohlman, T.~David, R.~Salem, S.~Aigner, J.~Schmiedmayer, and
  R.~Folman.
\newblock {Model for Organized Current Patterns in Disordered Conductors}.
\newblock {\em Phys. Rev. B}, 77:201407(R), 2008.
\newblock \href{http://journals.aps.org/prb/pdf/10.1103/PhysRevB.77.201407}{\tt
  doi=10.1103/PhysRevB.77.201407}.

\bibitem{Folman}
R.~Folman, P.~Treutlein, and J.~Schmiedmayer.
\newblock {Atom Chip Fabrication}.
\newblock In J.~Reichel and V.~Vuleti\'c, editors, {\em {Atom Chips}},
  chapter~3, page~61. Wiley-VCH, Weinheim, Germany, 2011.
\newblock
  \href{http://onlinelibrary.wiley.com/doi/10.1002/9783527633357.ch3/summary}{\tt
  doi=10.1002/9783527633357.ch3}.

\bibitem{Reichel}
J.~Reichel.
\newblock {Trapping and Manipulating Atoms on Chips}.
\newblock In J.~Reichel and V.~Vuleti\'c, editors, {\em {Atom Chips}},
  chapter~2, page~33. Wiley-VCH, Weinheim, Germany, 2011.
\newblock
  \href{http://onlinelibrary.wiley.com/doi/10.1002/9783527633357.ch2/summary}{\tt
  doi=10.1002/9783527633357.ch2}.

\bibitem{JPB35-469}
S.~Kraft, A.~G\"unther, H.~Ott, D.~Wharam, C.~Zimmermann, and J.~Fort\'agh.
\newblock {Anomalous longitudinal magnetic field near the surface of copper
  conductors}.
\newblock {\em J. Phys. B}, 35:L469, 2002.
\newblock
  \href{http://iopscience.iop.org/0953-4075/35/21/102/pdf/0953-4075_35_21_102.pdf}{\tt
  doi=10.1088/0953-4075/35/21/102}.

\bibitem{PRL98-263201}
J.-B. Trebbia, C.~L.~Garrido Alzar, R.~Cornelussen, C.~I. Westbrook, and
  I.~Bouchoule.
\newblock {Roughness Suppression via Rapid Current Modulation on an Atom Chip}.
\newblock {\em Phys. Rev. Lett.}, 98:263201, 2007.
\newblock
  \href{http://journals.aps.org/prl/pdf/10.1103/PhysRevLett.98.263201}{\tt
  doi=10.1103/PhysRevLett.98.263201}.

\bibitem{PRA75-063406}
T.~Fernholz, R.~Gerritsma, P.~Kr\"uger, and R.~J.~C. Spreeuw.
\newblock {Dynamically controlled toroidal and ring-shaped magnetic traps}.
\newblock {\em Phys. Rev. A}, 75:063406, 2007.
\newblock \href{http://journals.aps.org/pra/pdf/10.1103/PhysRevA.75.063406}{\tt
  doi=10.1103/PhysRevA.75.063406}.

\bibitem{AdvAtMol48-263}
R.~Folman, P.~Kr\"uger, J.~Schmiedmayer, J.~Denschlag, and C.~Henkel.
\newblock {Microscopic atom optics: from wires to an atom chip}.
\newblock {\em Adv. At. Mol. Opt. Phys.}, 48:263, 2002.
\newblock
  \href{http://www.sciencedirect.com/science/article/pii/S1049250X02800118}{\tt
  doi=10.1016/S1049-250X(02)80011-8}.

\bibitem{APB74-469}
J.~Reichel.
\newblock {Microchip traps and Bose-Einstein condensation}.
\newblock {\em Appl. Phys. B}, 74:469, 2002.
\newblock \href{http://link.springer.com/article/10.1007/s003400200861}{\tt
  doi=10.1007/s003400200861}.

\bibitem{NuclPhys243-203}
G.~M.~Tino {\it et al.}
\newblock {Precision Gravity Tests with Atom Interferometry in Space}.
\newblock {\em Nucl. Phys. B, Proc. Suppl.}, 243-244:203, 2013.
\newblock
  \href{http://www.zarm.uni-bremen.de/uploads/tx_sibibtex/precision_gravity_tests_neu.pdf}{\tt
  doi=10.1016/j.nuclphysbps.2013.09.023}.

\bibitem{CAL}
{ISS-Cold Atom Laboratory}.
\newblock \url{http://coldatomlab.jpl.nasa.gov/}.

\bibitem{PRA84-033639}
R.~P. Kafle, D.~Z. Anderson, and A.~A. Zozulya.
\newblock {Analysis of a free oscillation atom interferometer}.
\newblock {\em Phys. Rev. A}, 84:033639, 2011.
\newblock \href{http://journals.aps.org/pra/pdf/10.1103/PhysRevA.84.033639}{\tt
  10.1103/PhysRevA.84.033639}.

\bibitem{PRA86-043613}
R.~H. Leonard and C.~A. Sackett.
\newblock {Effect of trap anharmonicity on a free-oscillation atom
  interferometer}.
\newblock {\em Phys. Rev. A}, 86:043613, 2012.
\newblock \href{http://journals.aps.org/pra/pdf/10.1103/PhysRevA.86.043613}{\tt
  doi=10.1103/PhysRevA.86.043613}.

\bibitem{PRL99-060402}
Y.~Japha, O.~Arzouan, Y.~Avishai, and R.~Folman.
\newblock {Using Time-Reversal Symmetry for Sensitive Incoherent Matter-Wave
  Sagnac Interferometry}.
\newblock {\em Phys. Rev. Lett.}, 99:060402, 2007.
\newblock
  \href{http://journals.aps.org/prl/pdf/10.1103/PhysRevLett.99.060402}{\tt
  doi=10.1103/PhysRevLett.99.060402}.

\bibitem{PRL99-173201}
S.~Wu, E.~Su, and M.~Prentiss.
\newblock {Demonstration of an Area-Enclosing Guided-Atom Interferometer for
  Rotation Sensing}.
\newblock {\em Phys. Rev. Lett.}, 99:173201, 2007.
\newblock \href{http://prl.aps.org/pdf/PRL/v99/i17/e173201}{\tt
  doi=10.1103/PhysRevLett.99.173201}.

\bibitem{NatComm4-2077}
T.~Berrada, S.~{van Frank}, R.~B\"ucker, T.~Schumm, J.-F. Schaff, and
  J.~Schmiedmayer.
\newblock {Integrated Mach-Zehnder interferometer for Bose-Einstein
  condensates}.
\newblock {\em Nature Commun.}, 4:2077, 2013.
\newblock
  \href{http://www.nature.com/ncomms/2013/130627/ncomms3077/pdf/ncomms3077.pdf}{\tt
  doi=10.1038/ncomms3077}.

\bibitem{PRA79-040304}
D.~Petrosyan, G.~Bensky, G.~Kurizki, I.~Mazets, J.~Majer, and J.~Schmiedmayer.
\newblock {Reversible state transfer between superconducting qubits and atomic
  ensembles}.
\newblock {\em Phys. Rev. A}, 79:040304(R), 2009.
\newblock \href{http://journals.aps.org/pra/pdf/10.1103/PhysRevA.79.040304}{\tt
  doi=10.1103/PhysRevA.79.040304}.

\bibitem{PRL103-043603}
J.~Verd\'u, H.~Zoubi, Ch. Koller, J.~Majer, H.~Ritsch, and J.~Schmiedmayer.
\newblock {Strong Magnetic Coupling of an Ultracold Gas to a Superconducting
  Waveguide Cavity}.
\newblock {\em Phys. Rev. Lett.}, 103:043603, 2009.
\newblock
  \href{http://journals.aps.org/prl/pdf/10.1103/PhysRevLett.103.043603}{\tt
  doi=10.1103/PhysRevLett.103.043603}.

\bibitem{JPB35-2383}
Y.~Japha and Y.~B. Band.
\newblock {Motion of a condensate in a shaken and vibrating harmonic trap}.
\newblock {\em J. Phys. B}, 35:2383, 2002.
\newblock
  \href{http://iopscience.iop.org/0953-4075/35/10/315/pdf/0953-4075_35_10_315.pdf}{\tt
  doi=10.1088/0953-4075/35/10/315}.

\bibitem{PRA72-033610}
D.~M. Harber, J.~M. Obrecht, J.~M. McGuirk, and E.~A. Cornell.
\newblock {Measurement of the Casimir-Polder force through center-of-mass
  oscillations of a Bose-Einstein condensate}.
\newblock {\em Phys. Rev. A}, 72:033610, 2005.
\newblock \href{http://journals.aps.org/pra/pdf/10.1103/PhysRevA.72.033610}{\tt
  doi=10.1103/PhysRevA.72.033610}.

\bibitem{Nolte}
D.~D. Nolte.
\newblock {The tangled tale of phase space}.
\newblock {\em Phys. Today}, 63(4):33, 2010.
\newblock
  \href{http://scitation.aip.org/content/aip/magazine/physicstoday/article/63/4/10.1063/1.3397041}{\tt
  doi=10.1063/1.3397041}.

\bibitem{Leonhardt}
U.~Leonhardt.
\newblock {\em {Measuring the Quantum State of Light}}.
\newblock Cambridge University Press, New York, 2005.
\newblock
  \href{http://www.cambridge.org/us/academic/subjects/physics/optics-optoelectronics-and-photonics/measuring-quantum-state-light}{\tt
  ISBN=978-0-521-02352-8}.

\bibitem{PRL87-050402}
A.~I. Lvovsky, H.~Hansen, T.~Aichele, O.~Benson, J.~Mlynek, and S.~Schiller.
\newblock {Quantum State Reconstruction of the Single-Photon Fock State}.
\newblock {\em Phys. Rev. Lett.}, 87:050402, 2001.
\newblock \href{http://prl.aps.org/pdf/PRL/v87/i5/e050402}{\tt
  doi=10.1103/PhysRevLett.87.050402}.

\bibitem{PRL77-4281}
D.~Leibfried, D.~M. Meekhof, B.~E. King, C.~Monroe, W.~M. Itano, and D.~J.
  Wineland.
\newblock {Experimental Determination of the Motional Quantum State of a
  Trapped Atom}.
\newblock {\em Phys. Rev. Lett.}, 77:4281, 1996.
\newblock
  \href{https://journals.aps.org/prl/pdf/10.1103/PhysRevLett.77.4281}{\tt
  doi=10.1103/PhysRevLett.77.4281}.

\bibitem{Nature464-1170}
M.~F. Riedel, P.~B\"ohi, Y.~Li, T.~W. H\"ansch, A.~Sinatra, and P.~Treutlein.
\newblock {Atom-chip-based generation of entanglement for quantum metrology}.
\newblock {\em Nature (London)}, 464:1170, 2010.
\newblock
  \href{http://www.nature.com/nature/journal/v464/n7292/pdf/nature08988.pdf}{\tt
  doi=10.1038/nature08988}.

\bibitem{Nature386-150}
Ch. Kurtsiefer, T.~Pfau, and J.~Mlynek.
\newblock {Measurement of the Wigner function of an ensemble of helium atoms}.
\newblock {\em Nature (London)}, 386:150, 1997.
\newblock
  \href{http://www.nature.com/nature/journal/v386/n6621/pdf/386150a0.pdf}{\tt
  doi=10.1038/386150a0}.

\bibitem{JMO44-2551}
T.~Pfau and Ch. Kurtsiefer.
\newblock {Partial reconstruction of the motional Wigner function of an
  ensemble of helium atoms}.
\newblock {\em J. Mod. Optic.}, 44:2551, 1997.
\newblock
  \href{http://www.tandfonline.com/doi/pdf/10.1080/09500349708231900}{\tt
  doi=10.1080/09500349708231900}.

\bibitem{NJP5-71}
F.~S. Cataliotti, L.~Fallani, F.~Ferlaino, C.~Fort, P.~Maddaloni, and
  M.~Inguscio.
\newblock {Superfluid current disruption in a chain of weakly coupled
  Bose-Einstein condensates}.
\newblock {\em New J. Phys.}, 5:71, 2003.
\newblock
  \href{http://iopscience.iop.org/1367-2630/5/1/371/pdf/1367-2630_5_1_371.pdf}{\tt
  doi=10.1088/1367-2630/5/1/371}.

\bibitem{JPB43-155002}
J.~J.~P. {van Es}, P.~Wicke, A.~H. {van Amerongen}, C.~R\'etif, S.~Whitlock,
  and N.~J. {van Druten}.
\newblock {Box traps on an atom chip for one-dimensional quantum gases}.
\newblock {\em J. Phys. B}, 43:155002, 2010.
\newblock
  \href{http://iopscience.iop.org/0953-4075/43/15/155002/pdf/0953-4075_43_15_155002.pdf}{\tt
  doi=10.1088/0953-4075/43/15/155002}.

\bibitem{PRA81-063415}
A.~Bertoldi and L.~Ricci.
\newblock {Dynamics of a cold atom cloud in an anharmonic trap}.
\newblock {\em Phys. Rev. A}, 81:063415, 2010.
\newblock \href{http://journals.aps.org/pra/pdf/10.1103/PhysRevA.81.063415}{\tt
  doi=10.1103/PhysRevA.81.063415}.

\bibitem{PRA88-043406}
I.~{Llorente Garc\'ia}, B.~Darqui\'e, C.~D.~J. Sinclair, E.~A. Curtis,
  M.~Tachikawa, J.~J. Hudson, and E.~A. Hinds.
\newblock {Shaking-induced dynamics of cold atoms in magnetic traps}.
\newblock {\em Phys. Rev. A}, 88:043406, 2013.
\newblock \href{http://journals.aps.org/pra/pdf/10.1103/PhysRevA.88.043406}{\tt
  doi=10.1103/PhysRevA.88.043406}.

\bibitem{PRA78-025602}
A.~del Campo, V.~I. Man’ko, and G.~Marmo.
\newblock {Symplectic tomography of ultracold gases in tight waveguides}.
\newblock {\em Phys. Rev. A}, 78:025602, 2008.
\newblock \href{http://journals.aps.org/pra/pdf/10.1103/PhysRevA.78.025602}{\tt
  doi=10.1103/PhysRevA.78.025602}.

\bibitem{EPL59-694}
J.~\v{R}eh\'a\v{c}ek, Z.~Hradil, M.~Zawisky, W.~Treimer, and M.~Strobl.
\newblock {Maximum-likelihood absorption tomography}.
\newblock {\em Europhys. Lett.}, 59:694, 2002.
\newblock
  \href{http://iopscience.iop.org/0295-5075/59/5/694/pdf/0295-5075_59_5_694.pdf}{\tt
  doi=10.1209/epl/i2002-00181-4}.

\bibitem{PRL78-2088}
H.~Ammann and N.~Christensen.
\newblock {Delta Kick Cooling: A New Method for Cooling Atoms}.
\newblock {\em Phys. Rev. Lett.}, 78:2088, 1997.
\newblock
  \href{http://journals.aps.org/prl/pdf/10.1103/PhysRevLett.78.2088}{\tt
  doi=10.1103/PhysRevLett.78.2088}.

\bibitem{PRL85-1795}
S.~L. Cornish, N.~R. Claussen, J.~L. Roberts, E.~A. Cornell, and C.~E. Wieman.
\newblock {Stable $\rm^{85}Rb$ Bose-Einstein Condensates with Widely Tunable
  Interactions}.
\newblock {\em Phys. Rev. Lett.}, 85:1795, 2000.
\newblock
  \href{http://journals.aps.org/prl/pdf/10.1103/PhysRevLett.85.1795}{\tt
  doi=10.1103/PhysRevLett.85.1795}.

\bibitem{PRA67-060701}
N.~R. Claussen, S.~J. J. M.~F. Kokkelmans, S.~T. Thompson, E.~A. Donley,
  E.~Hodby, and C.~E. Wieman.
\newblock {Very-high-precision bound-state spectroscopy near a $\rm^{85}Rb$
  Feshbach resonance}.
\newblock {\em Phys. Rev. A}, 67:060701, 2003.
\newblock \href{http://journals.aps.org/pra/pdf/10.1103/PhysRevA.67.060701}{\tt
  doi=10.1103/PhysRevA.67.060701}.

\bibitem{RSI81-063103}
P.~A. Altin, N.~P. Robins, D.~D\"oring, J.~E. Debs, R.~Poldy, C.~Figl, and
  J.~D. Close.
\newblock {$\rm^{85}Rb$ tunable-interaction Bose-Einstein condensate machine}.
\newblock {\em Rev. Sci. Instrum.}, 81:063103, 2010.
\newblock
  \href{http://scitation.aip.org/content/aip/journal/rsi/81/6/10.1063/1.3430538}{\tt
  doi=10.1063/1.3430538}.

\bibitem{NatComm4-2424}
S.~Machluf, Y.~Japha, and R.~Folman.
\newblock {Coherent Stern-Gerlach momentum splitting on an atom chip}.
\newblock {\em Nature Commun.}, 4:2424, 2013.
\newblock
  \href{http://www.nature.com/ncomms/2013/130909/ncomms3424/full/ncomms3424.html}{\tt
  doi=10.1038/ncomms3424}.

\bibitem{Nature440-900}
T.~Kinoshita, T.~Wenger, and D.~S. Weiss.
\newblock {A quantum Newton’s cradle}.
\newblock {\em Nature (London)}, 440:900, 2006.
\newblock
  \href{http://www.nature.com/nature/journal/v440/n7086/pdf/nature04693.pdf}{\tt
  doi=10.1038/nature04693}.

\bibitem{NJP12-055023}
I.~E. Mazets and J.~Schmiedmayer.
\newblock {Thermalization in a quasi-one-dimensional ultracold bosonic gas}.
\newblock {\em New J. Phys.}, 12:055023, 2010.
\newblock
  \href{http://iopscience.iop.org/1367-2630/12/5/055023/pdf/1367-2630_12_5_055023.pdf}{\tt
  doi=10.1088/1367-2630/12/5/055023}.

\bibitem{JOptB5-420}
C.-Y. Wong.
\newblock {Explicit solution of the time evolution of the Wigner function}.
\newblock {\em J. Opt. B}, 5:S420, 2003.
\newblock
  \href{http://iopscience.iop.org/1464-4266/5/3/381/pdf/1464-4266_5_3_381.pdf}{\tt
  doi=10.1088/1464-4266/5/3/381}.

\bibitem{PRA81-065602}
F.~Impens and D.~Gu\'ery-Odelin.
\newblock {Classical phase-space approach for coherent matter waves}.
\newblock {\em Phys. Rev. A}, 81:065602, 2010.
\newblock \href{http://journals.aps.org/pra/pdf/10.1103/PhysRevA.81.065602}{\tt
  10.1103/PhysRevA.81.065602}.

\bibitem{PRA55-2109}
H.~Wallis, A.~R\"ohrl, M.~Naraschewski, and A.~Schenzle.
\newblock {Phase-space dynamics of Bose condensates: Interference versus
  interaction}.
\newblock {\em Phys. Rev. A}, 55:2109, 1997.
\newblock \href{http://journals.aps.org/pra/pdf/10.1103/PhysRevA.55.2109}{\tt
  10.1103/PhysRevA.55.2109}.

\bibitem{PRA41-3436}
A.~Stern, Y.~Aharonov, and Y.~Imry.
\newblock {Phase uncertainty and loss of interference: A general picture}.
\newblock {\em Phys. Rev. A}, 41:3436, 1990.
\newblock \href{http://journals.aps.org/pra/pdf/10.1103/PhysRevA.41.3436}{\tt
  10.1103/PhysRevA.41.3436}.

\end{thebibliography}

\end{document}